OPEN

# A downsampling strategy to assess the predictive value of radiomic features

Anne-Sophie Dirand*, Frédérique Frouin & Irène Buvat

**Many studies are devoted to the design of radiomic models for a prediction task. When no effective model is found, it is often difficult to know whether the radiomic features do not include information relevant to the task or because of insufficient data. We propose a downsampling method to answer that question when considering a classification task into two groups. Using two large patient cohorts, several experimental configurations involving different numbers of patients were created. Univariate or multivariate radiomic models were designed from each configuration. Their performance as reflected by the Youden index (YI) and Area Under the receiver operating characteristic Curve (AUC) was compared to the stable performance obtained with the highest number of patients. A downsampling method is described to predict the YI and AUC achievable with a large number of patients. Using the multivariate models involving machine learning, YI and AUC increased with the number of patients while they decreased for univariate models. The downsampling method better estimated YI and AUC obtained with the largest number of patients than the YI and AUC obtained using the number of available patients and identifies the lack of information relevant to the classification task when no such information exists.**

Radiomics is an expanding domain based on the extraction of mineable high dimensional data from medical images[1]. So-called radiomic features extracted from the images are supposed to carry information about tumor biology and patient prognosis that could assist clinical decision making. A wide range of models involving radiomic features have been proposed for prediction tasks, such as predicting the response to a treatment or the patient progression free survival[2]. Many statistical methods such as Receiver Operating Characteristic (ROC) analysis can give insights on whether radiomic features have some predictive values. Predictive radiomic models can also be designed using Machine Learning (ML) approaches, involving Logistic Regression (LR), Linear Discriminant Analysis, Support Vector Machine (SVM), or Random Forests, among others. Deep learning is also increasingly investigated to build predictive or classification models. There is currently no consensus on which approach should be preferred to design a radiomic model, and different ML methods can yield significantly different performance[3]. A Pubmed search with ((CT radiomics) OR (PET radiomics) OR (MRI radiomics)) AND ((“2017”[Date - Publication]) OR (“2018”[Date - Publication])) criteria highlighted 452 articles from which 189 were about developing radiomic models for a classification task while the others reported methodological developments or addressed survival prediction tasks. Among these 189 articles, 158 used ML methods, 9 used univariate ROC analysis and 22 used deep learning methods.

Once a model is created, its robustness and performance on data acquired in different centers are key elements for subsequent clinical translation[4]. Ideally, independent datasets processed by independent investigators should be used to validate the model in different settings. At best, in 93 of the 189 above reported publications, the model was evaluated on a test data set that was not used to build the model. From these 93 articles, only 19 studies used a test set obtained from a different cohort.

The number of enrolled patients (or cases) for designing, validating and testing a radiomic model is highly variable across studies. In the 189 articles, the median was 96 patients (range: 13–1294) in the training and validation sets. For the 93 articles including a test set, the median was 50 patients in the test set (range: 1–237). The repartition of the patients in the training sub-groups to be distinguished by the model also varied substantially. For the 161 articles from which the repartition was described, 66% of the training sets were substantially unbalanced, i.e. with less than 40% of the patients in one of the two classes.

Imagerie Moléculaire In Vivo, CEA-SHJF, Inserm, CNRS, Université Paris-Sud, Université Paris-Saclay, Orsay, France. *email: dirandannesophie@gmail.com

This large variability in study design adds complexity to compare results reported in different articles. In particular, when a study fails to identify a predictive model with satisfactory performance, it is often difficult to determine whether the radiomic features do not include relevant data for the classification task, whether an inadequate model has been used, or whether the number of patients was not sufficient.

To the best of our knowledge, the sample size required to build robust classification models in the context of radiomics has not been investigated yet, although this question has been addressed in the non-imaging field[5]. The goal of our study was therefore twofold: first to investigate the reliability of the predicted performance of classification models as a function of the experimental conditions and second to design a method that determines whether data contain predictive information for a given task from a limited dataset. The downsampling method proposed for that second purpose does not aim at identifying the best classification model but at estimating which classification performance can be expected. Our work was based on the study of two large cohorts of patients from the literature and for which features extracted from 18F-FDG-PET/CT and fine needle aspiration images yielded effective predictive models with a Youden index (=Sensitivity + Specificity − 1) greater than 0.70 and Area Under the receiver operating characteristic Curve AUC greater than 0.90. From these cohorts, we first created various experimental set-ups, involving different sample sizes and various statistical and machine learning methods, to determine the conditions required to identify the presence of a valuable predictive information from the data. From these observations, we designed an original approach based on a downsampling strategy to predict the performance to be expected when enough patients are available.

## Materials and Methods

**Cohorts.** Cohort 1 was derived from[6], in which the authors investigated the ability of 18F-FDG-PET/CT based radiomics features to predict whether a lung lesion is a primary lung lesion (PL) or a metastasis (MT). The reported study was approved by the institutional Ethic Committee. The cohort included 414 PET lesions, including 303 PL and 111 MT. For each lesion, 43 radiomic features were calculated[7] from the PET images and corresponding values were provided as Supplemental Data. Linear Discriminant Analysis classifiers (a direct one and a backward one) were designed from 307 patients of the cohort and the model performance was evaluated using 100 patients. This process was repeated 100 times to study the robustness of the results.

Cohort 2 was derived from[8], where the task was to distinguish between malignant (MM) and Benign Fibrocystic Breast Mass (BM) lesions based on nuclear features extracted from breast fine needle aspiration (FNA) images. The cohort was composed of 569 FNA digital images, including 212 MM and 357 BM. For each image, 30 features, namely the mean values, mean of the three largest values and standard errors of 10 parameters calculated for each nucleus, were extracted and made publicly available. The model to distinguish between BM and MM was built using a Logistic Regression approach with a 10-fold cross-validation. The process was repeated 100 times and no additional test of the model was performed.

**Experimental set-ups.** For the two cohorts, we designed various set-ups making use of only part of the cohorts to investigate how the estimated model and classification performance evolve as a function of the data that are available. We also created two synthetic data sets from each of these cohorts (referred to as synthetic cohort 1 or 2), with no predictive information. To that end, from cohort 1, each patient with its associated radiomic features was randomly assigned to the PL or MT group, respecting the fraction of PL to MT observed in the real cohort 1. Similarly, in cohort 2, each patient with its associated FNA features was randomly assigned to the MM or BM group, with the same fraction of MM to BM patients as in cohort 2, to create synthetic cohort 2. Doing so, we did not expect to find any model that could predict to which group a patient was assigned based on its associated image-based features. The performance of different classification methods on these fake data was also studied.

For each cohort, each set-up and each type of model building approach (ROC univariate analysis, SVM or LR with or without dimensionality reduction and least absolute shrinkage and selection operator linear model (LASSO)), the data set was randomly split into two subsets. The first one included the training set (TRS) and the validation set (VS) and was used to develop and fine tune the model. The second one, called test set (TES), was used to independently assess the performance of the model. The performance of each classification model was measured using the Youden index (YI) and the Area Under the Curve. For cohort 1 and synthetic cohort 1, 14 set-ups (S1 to S14) were investigated, varying in the total number of lesions included and in the ratio between PL and MT lesions (Table 1). For the unbalanced cases (S1 to S9), the number of patients started from 40 (30 PL and 10 MT) and was incremented from one set-up to the next by 40 (30 PL and 10 MT), corresponding to the ratio observed in the whole cohort, up to a total number of 360 (270 PL and 90 MT). For the balanced cases (S10 to S14), the number of patients started from 60 (30 PL and 30 MT) and was incremented from one set-up to another by 30 (15 PL and 15 MT) up to a total number of 180 (90 PL and 90 MT). The test set was always composed of 40 patients (30 PL and 10 MT).

For cohort 2 and synthetic cohort 2, 18 set-ups (S1 to S18) were investigated, varying in the total number of images from FNA and in the ratio between MM and BM (Supplementary Table 1). For the unbalanced cases (S1 to S9), the number of patients started from 27 (17 BM and 10 MM) and was incremented from one set-up to another up to a total number of 216 (136 BM and 80 MM), the last set-up S9 included 270 BM and 160 MM. For the balanced cases (S10 to S18), the number of patients started from 20 (10 BM and 10 MB) and was incremented from one set-up to another by 20 (10 BM and 10 MM) up to a total number of 160 (80 BM and 80 MM) and an extra set-up S18 included 320 cases (160 BM and 160 MM). The test set was always composed of 113 patients (71 BM and 42 MM).

For each set-up, a different random split of the whole data set was repeated 50 times, meaning that training, validation and test datasets changed in each of the 50 runs to study the stability of the results.

| Set-up | Training set + Validation set (TRS + VS) | | External testing (ETS) | | Number of folds for the SKFCV |
|---|---|---|---|---|---|
| | PL | MT | PL | MT | |
| S1 | 30 | 10 | 30 | 10 | 2 |
| S2 | 60 | 20 | 30 | 10 | 2 |
| S3 | 90 | 30 | 30 | 10 | 3 |
| S4 | 120 | 40 | 30 | 10 | 3 |
| S5 | 150 | 50 | 30 | 10 | 3 |
| S6 | 180 | 60 | 30 | 10 | 3 |
| S7 | 210 | 70 | 30 | 10 | 3 |
| S8 | 240 | 80 | 30 | 10 | 3 |
| S9 | 270 | 90 | 30 | 10 | 3 |
| **S10** | **30** | **30** | **30** | **10** | **3** |
| **S11** | **45** | **45** | **30** | **10** | **4** |
| **S12** | **60** | **60** | **30** | **10** | **6** |
| **S13** | **75** | **75** | **30** | **10** | **7** |
| **S14** | **90** | **90** | **30** | **10** | **9** |

**Table 1.** Number of patients for each set-up of cohort 1 and synthetic cohort 1. Balanced situations in bold characters.

**Sensitivity of the univariate analysis to the number of patients.** For the four cohorts, the performance of the univariate classification models was assessed using the area under the ROC curves and the Youden index given with its corresponding sensitivity (Se) and specificity (Sp) values. The optimal cut-off to separate the groups was defined as the cut-off value that maximized YI.

For each run (i) of each set-up, we performed a ROC study for each feature using the training and validation set (TRS + VS) to define the optimal cut-off as the one maximizing the YI obtained on (TRS + VS), called $Y_{i,ROC}^{TRS+VS}$. $Y_{i,ROC}^{TRS+VS}$ was compared to the Youden index obtained on the test set (TES), $Y_{i,ROC}^{TES}$, using that cut-off. The mean of the fifty $Y_{i,ROC}^{TRS+VS}$ and $Y_{i,ROC}^{TES}$ values were defined as $Y_{ROC}^{TRS+VS}$ and $Y_{ROC}^{TES}$ respectively. The behavior of $Y_{ROC}^{TRS+VS}$ depending on the number of patients was studied as well as the difference between the predicted performance $Y_{ROC}^{TRS+VS}$ and the actual performance observed on test data $Y_{ROC}^{TES}$ using the cut-off determined from the training data. The process was repeated independently for each radiomic feature and is illustrated in Fig. 1a. The AUC computed from TRS + VS were also given.

**Sensitivity of the multivariate analysis to the number of patients.** For the four cohorts, we studied three types of classifier implemented in scikit learn toolbox[9]: LR, SVM with a linear kernel, and LASSO. For these 3 classifiers, the following parameters were tuned: inverse of regularization strength for LR, penalty parameter for SVM and constant that multiplies the L1 term for LASSO. LR uses a logistic function to model a binary dependent variable, SVM identifies the hyper-plane that best differentiates the two classes and LASSO is a regression analysis method that performs a variable selection as well as a regularization. LR and SVM were investigated either on their own, or in association with a dimensionality reduction method. Three dimensionality reduction methods were studied: Recursive Feature Elimination (RFE), Principal Component Analysis (PCA) and $ROC_{fr}$ analysis. RFE selects features recursively according to the coefficients describing their importance. PCA enables dimension reduction by transforming a number of correlated variables into a smaller subset of orthogonal principal components. $ROC_{fr}$ consists in using the ROC univariate analysis to select features yielding a YI greater than a specific threshold for each cohort.

For all methods, the imbalance between the two groups, if any, was taken into account during the training, assigning weights to the groups. For the sake of clarity, each ML method is furthered referred to $C_m$, m varying from 1 to 9: $C_1$ for LR, $C_2$ for RFE-LR, $C_3$ for PCA-LR, $C_4$ for $ROC_{fr}$-LR, $C_5$ for SVM, $C_6$ for RFE-SVM, $C_7$ for PCA-SVM, $C_8$ for $ROC_{fr}$-SVM and $C_9$ for LASSO. To characterize the performance of the classification models, areas under the ROC curves and Youden index with its corresponding sensitivity and specificity values were used. To design the classification model, the parameters of the classifiers were tuned using a Stratified K-Fold Cross Validation (SKFCV) with K folds. The numbers of folds for each set-up and each cohort are given in Table 1 and Supplementary Table 1. For each run i and each ML method $C_m$, the model was developed on TRS and then applied on VS. For each fold, each feature was normalized with a z-score calculated from the mean and standard deviation of TRS and then VS and TES were normalized using these mean and standard deviation. The selected model was the one yielding the highest YI $Y_{i,C_m}^{VS-K}$ among the K VS of the fold. This model was finally applied to TES to estimate the corresponding $Y_{i,C_m}^{TES}$. $Y_{i,C_m}^{VS}$ was calculated as the mean of the K $Y_{i,C_m}^{VS-K}$ (with its associated standard deviation $SY_{i,C_m}^{VS}$) to be compared to $Y_{i,C_m}^{TES}$. The mean of the fifty $Y_{i,C_m}^{VS}$ (respectively $Y_{i,C_m}^{TES}$) was noted $Y_{C_m}^{VS}$ (respectively $Y_{C_m}^{TES}$) with its associated standard deviation $SY_{C_m}^{VS}$ (respectively $SY_{C_m}^{TES}$). The mean difference and absolute difference between $Y_{i,C_m}^{VS}$ and $Y_{i,C_m}^{TES}$ were also calculated. The process is illustrated in Fig. 1b.

**Prediction of the classification model performance - Univariate models.** To predict the performance in terms of YI of a model from a reduced number of cases, we determined how the estimated $Y_{ROC}^{TRS+VS}$ varied with the number of cases and extrapolated from that trend the YI to expect when enough cases are available. Let us assume that a set of N cases is available. From these N cases, we generated P subsets ($sS_1$, $sS_2$, …, $sS_P$) by downsampling the N cases with a step of S cases. Subsample $sS_p$ included $N_p = N - (p-1)S$ cases, $sS_1$ including all the N available cases. We therefore obtained $Y_{i,ROC}^{TRS+VS}(N_p)$ for each subsample $sS_p$. This was repeated 50 times and the 50 $Y_{i,ROC}^{TRS+VS}(N_p)$ values for each subsample were averaged to yield $Y_{ROC}^{TRS+VS}(N_p)$. The standard deviation $SY_{ROC}^{TRS+VS}(N_p)$ over the 50 runs was also calculated.

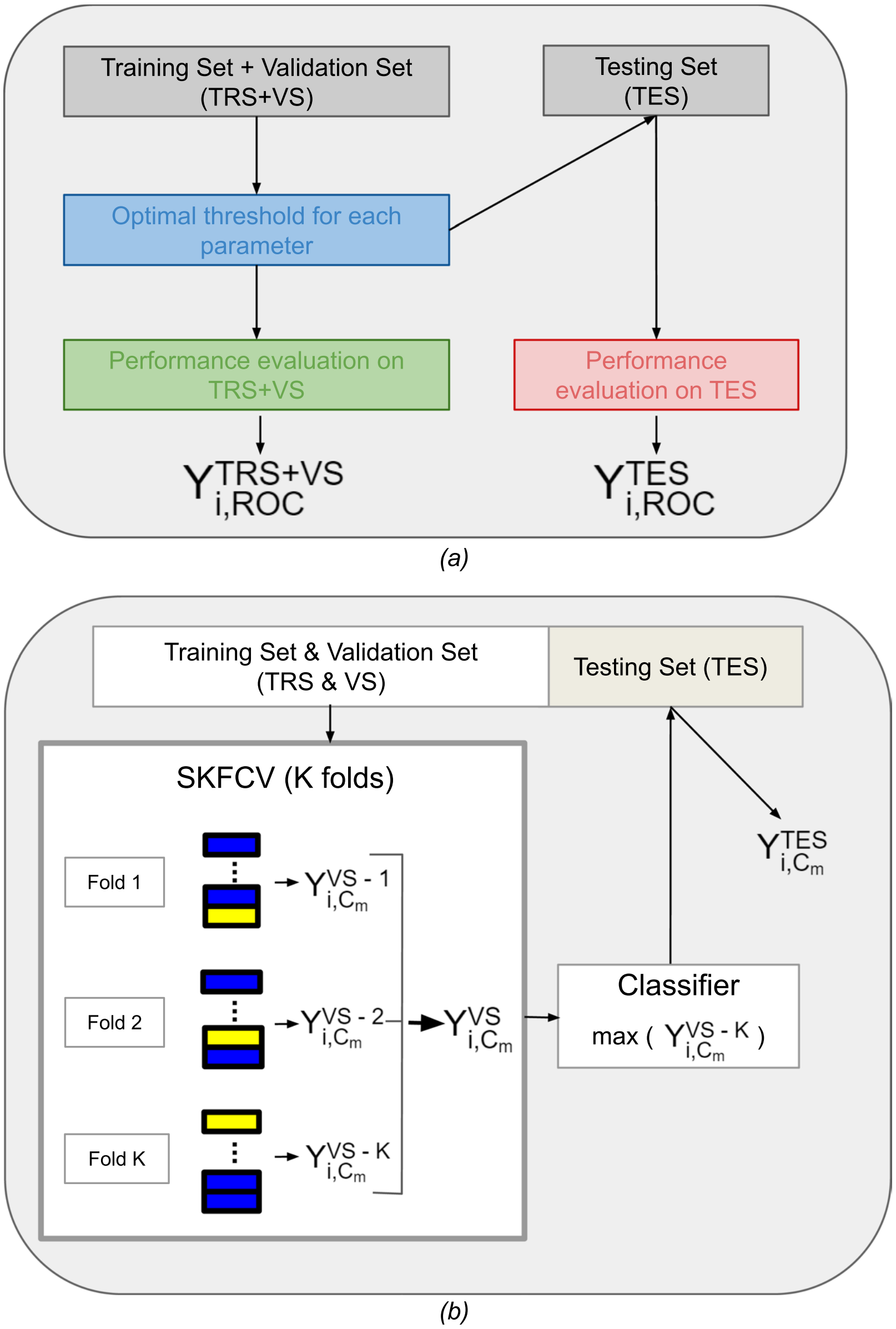

**Figure 1.** (**a**) Pipeline of the ROC analysis. Y is the Youden index and i the considered run. (**b**) Pipeline of the machine learning process. Y is the Youden index, i the considered run and $C_m$ the considered classifier.

For each feature, $Y_{ROC}^{TRS+VS}(N_p)$ got stable from a certain number of patients $N_p$ and a plateau was reached. We empirically defined stability when $Y_{ROC}^{TRS+VS}(N_p + S)$- $Y_{ROC}^{TRS+VS}(N_p)$ did not change by more than 0.01 between 2 consecutive subsets composed of $N_p$ and $N_p + S$ cases.

Using the P subsamples $sS_p$, we plotted $SY_{ROC}^{TRS+VS}$ as a function of the number of patients $N_p$. The resulting points were fitted using a linear model (Eq. 1):

$$SY_{ROC}^{TRS+VS}(N_p) = a_m - b_m.\ N_p \quad (1)$$

The intersection of the fitted line with $SY_{ROC}^{TRS+VS} = 0$ gave a number of patients $N_c$. We then fitted the $Y_{ROC}^{TRS+VS}(N_p)$ curve using a logarithmic model where each point of the plot is weighted by the inverse of the standard deviation $SY_{ROC}^{TRS+VS}(N_p)$ associated with $Y_{ROC}^{TRS+VS}(N_p)$ (Eq. 2):

$$Y_{ROC}^{TRS+VS}(N_p) = c_m + d_m.\ \log(N_p) \text{ with } d_m < 0 \tag{2}$$

The expected performance at convergence ($Yec_{ROC}(N)$) was finally obtained by extrapolating $Y_{ROC}^{TRS+VS}(N_p)$ for $N_p = N_c$ using Eq. 2.

Note that $N_c$ is not necessarily the number of patients needed to reach stability but rather a number of patients needed to predict the performance of the classifier. The ability to predict AUC was also investigated using the exact same process and the same fits, ie by replacing the Youden figure of merit by the AUC figure of merit.

**Prediction of the classification model performance - Multivariate models.** The same process was applied to the results of each of the 9 ML classifiers. The stability criterion was the same as the one for univariate analysis. Since $SY_{C_m}^{VS}$ and $Y_{C_m}^{VS}$ changes as a function of the number of patients were not the same as $SY_{ROC}^{TRS+VS}$ and $Y_{ROC}^{TRS+VS}$, the fit used to estimate $N_c$ was different. A logarithmic model was used to estimate the number of patients $N_c$ from the variation of $SY_{C_m}^{VS}(N_p)$ as a function of $N_p$ (Eq. 3):

$$SY_{C_m}^{VS}(N_p) = a'_m - b'_m.\ \log(N_p) \tag{3}$$

The intersection of the fit with $SY_{C_m}^{VS} = 0$ yielded $N_c$. The $Y_{C_m}^{VS}(N_p)$ points were fitted using a logarithmic model where each point of the plot was weighted by the inverse of the standard deviation $SY_{C_m}^{VS}(N_p)$ associated with $Y_{C_m}^{VS}(N_p)$ (Eq. 4)

$$Y_{C_m}^{VS}(N_p) = c_m + d_m.\ \log(N_p) \text{ with } d_m > 0 \tag{4}$$

Similar to the univariate analysis, the expected performance at convergence $Yec_{Cm}(N)$ was defined by extrapolating $Y_{C_m}^{VS}(N_p)$ for $N_p = N_c$.

The ability to predict AUC was also investigated using the exact same process and the same fits, ie by replacing the Youden figure of merit by the AUC figure of merit.

**Evaluation of the quality of the prediction - Univariate models.** To assess the accuracy of the predicted YI when using univariate analysis, we tested 17 experimental conditions for each feature with an unbalanced number of patients (from N = 36 to 164) for the real data of cohort 1 and for synthetic cohort 1. N starts from 36 to have at least 3 points to perform the fits (S = 8 patients). Similarly, we tested 19 experimental conditions for each feature with unbalanced number of patients (from N=19 to 73) for real cohort 2 and for synthetic cohort 2. A quality factor QF1 (Eq. 5):

$$\text{QF1} = |Yec_{ROC}(N) - Y_{ROC}^{TRS+VS}(N_{tot})| \tag{5}$$

was computed to determine how close the prediction $Yec_{ROC}(N)$ was from $Y_{ROC}^{TRS+VS}(N_{tot})$ obtained with the largest number of patients $N_{tot}$ ($N_{tot} = 414$ for cohort 1 and 569 for cohort 2) when stability was reached. As a comparison, a second quality factor QF2 (Eq. 6):

$$\text{QF2} = |Y_{ROC}^{TRS+VS}(N) - Y_{ROC}^{TRS+VS}(N_{tot})| \tag{6}$$

was computed to determine how close the performance obtained with the N patients $Y_{ROC}^{TRS+VS}(N)$ but without downsampling was from $Y_{ROC}^{TRS+VS}(N_{tot})$, as $Y_{ROC}^{TRS+VS}(N)$ is usually used as an estimate of the expected performance. Having QF1 < QF2 shows the usefulness of the downsampling approach and the closer QF1 to 0, the more accurate the prediction.

**Evaluation of the quality of the prediction - Multivariate models.** With ML approaches, we tested 24 experimental conditions with an unbalanced number of patients from cohort 1 differing by the number of patients N (from 36 to 220) for the 9 classifiers. To study the behavior of the downsampling method when no predictive model exists, 24 experimental conditions with an unbalanced number of patients for 5 classifiers (the previous ones except $C_2$, $C_4$, $C_6$ and $C_8$) were tested using synthetic cohort 1 that had no predictive information. The downsampling step was set to S = 8. $C_2$, $C_4$, $C_6$ and $C_8$ were not used since these techniques are based on an explicit selection of features, which is not relevant when there is no feature useful for the classification. Similarly, we studied 19 experimental conditions with an unbalanced number of patients for cohort 2, differing by the number of patients N (from 13 to 73) for the 9 classifiers. Furthermore, 19 experimental conditions with an unbalanced number of patients for 5 classifiers (the previous ones except $C_2$, $C_4$, $C_6$ and $C_8$) were tested using synthetic cohort 2. The downsampling step was set to S = 3. For these experimental conditions, we predicted $Yec_{C_m}(N)$ as described above.

The reliability of $Yec_{C_m}(N)$ was assessed using metrics similar to the ones defined for the univariate analysis with QF1′ (Eq. 7) and QF2′ (Eq. 8):

$$\text{QF1}' = \left|Yec_{C_m}(N) - Y_{C_m}^{VS}(N_{tot})\right| \tag{7}$$

$$\text{QF2}' = |Y_{C_m}^{VS}(N) - Y_{C_m}^{VS}(N_{tot})| \tag{8}$$

We also looked at the proportion of predicted AUC that were within 5% of the AUC obtained on the validation cohort when using all patients (414 for cohort 1 and 569 for cohort 2).

**Use case.** To illustrate the practical application of the proposed downsampling approach, we created a set-up from cohort 1 in which the data from only 88 patients would be available to build a LASSO model ($C_9$), as LASSO is often used in radiomics. We downsampled from 84 patients drawn out of the 88 with a downsampling step S of 8 patients to create a 9-points curve that was then fitted to get $Yec_{C_9}(84)$. We also calculated $Y_{C_9}^{VS}(88)$ which is usually what is performed to estimate the performance of the model based on an 88 patient sample. We then compared $Yec_{C_9}(84)$ and $Y_{C_9}^{VS}(88)$ to $Y_{C_9}^{VS}(414)$ obtained using a LASSO model built from the whole cohort of 414 patients. We repeated exactly the same analysis using AUC as a figure of merit.

## Results

For cohort 1[6], the original study reported a YI of 0.73 ± 0.06 (Se = 0.89 ± 0.02, Sp = 0.84 ± 0.04) and AUC = 0.92 ± 0.02 for the direct LDA approach and 0.74 ± 0.06 (Se = 0.89 ± 0.02, Sp = 0.85 ± 0.04) and AUC = 0.91 ± 0.02 for the backward LDA approach. For cohort 2[8], a YI of 0.92 ± 0.01 (Se = 0.94 ± 0.00 and Sp = 0.98 ± 0.00) was reported with a logistic regression approach.

For the sake of conciseness, the results are detailed for cohort 1 and only summarized for cohort 2, with all detailed results for cohort 2 provided as "Supplemental Data".

**Sensitivity of the univariate analysis to the number of patients.** The best performing feature that reached stability was SUVmin for cohort 1 with $Y_{ROC}^{TRS+VS} \pm SY_{ROC}^{TRS+VS} = 0.50 \pm 0.02$ (Se = 0.71 ± 0.11, Sp = 0.81 ± 0.05) and AUC = 0.79 ± 0.01. The YI obtained on TRS + VS ($Y_{ROC}^{TRS+VS}$) and on TES ($Y_{ROC}^{TES}$) as a function of the experimental conditions as well as the AUC are shown in Fig. 2 for cohort 1 and synthetic cohort 1 for $SUV_{min}$. Figures 2a,c show that stability is reached from S5, namely from 200 patients. Figures 2b,d show the results when using synthetic cohort 1 with no predictive information. In cohort 1 (Supplementary Table S2a), the number of features yielding $Y_{ROC}^{TRS+VS}$ greater than 0.20 decreased as the number of patients increased while the number of features leading to a $Y_{ROC}^{TES}$ greater than 0.20 increased with more patients. The ratio of these 2 numbers, which reflects the ability of the training process to select features that would remain selected in TES, increased with the number of patients. The trend was the same for both cohorts (Supplementary Table S2b for cohort 2). In cohort 2, a cut-off of $Y_{ROC}^{TRS+VS}$ greater than 0.75 was used to identify the most predictive features because almost all features yielded a $Y_{ROC}^{TRS+VS}$ greater than 0.20. The best performing feature in cohort 2 was the worst perimeter (mean of the 3 largest nuclei perimeters in the image) with a $Y_{ROC}^{TRS+VS} \pm SY_{ROC}^{TRS+VS}$ of 0.84 ± 0.01 (Se = 0.91 ± 0.02, Sp = 0.92 ± 0.02) and AUC = 0.98 ± 0.00.

When using all patients, 2 out of 43 features for cohort 1 and 16 out of 30 features for cohort 2 had a YI ≥ 0.5. AUC were equal to or greater than 0.75 for 2 out 43 features for cohort 1 and 20 out of 30 features for cohort 2.

When the data do not contain predictive information, $Y_{ROC}^{TRS+VS}$ is close to 0 (Fig. 2b for synthetic cohort 1 and Supplementary Fig. S1b for synthetic cohort 2). As expected, AUC is close to 0.5 (Fig. 2d for synthetic cohort 1 and Supplementary Fig. S1d for synthetic cohort 2). The trends for the performance described above were the same for the balanced set-ups (Fig. 2a,c for cohort 1 and Supplementary Fig. S1a,c for cohort 2). The same conclusions were drawn concerning the number of features selected on TRS + VS and on TES (Supplementary Table S2a for cohort 1 and Supplementary Table S2b for cohort 2).

**Sensitivity of the multivariate analysis to the number of patients.** The $Y_{C_m}^{VS}$ for $C_5$ are shown in Fig. 3a (Fig. 4a for $C_6$) for the 14 experimental conditions. The absolute differences between $Y_{C_m}^{VS}$ and $Y_{C_m}^{TES}$ are shown in Fig. 3b for $C_5$ and Fig. 4b for $C_6$ for cohort 1. In Figs. 3a and 4a, the closer to 1 $Y_{C_m}^{VS}$, the better and in Figs. 3b and 4b. the closer to 0 the absolute difference, the better. Figures corresponding to the other classifiers for cohorts 1 and 2 are given in Supplementary Figs. S2 to S17 for the Youden index. Results in terms of AUC are shown in Fig. 3c for $C_5$ and Fig. 4c for $C_6$ and absolute differences in AUC are shown in Fig. 3d for $C_5$ and Fig. 4d for $C_6$.

Table 2 and Supplementary Table S3 give the percentage of experiments for which $Y_{i,C_m}^{TES}$ was included in $Y_{i,C_m}^{VS} \pm SY_{i,C_m}^{VS}$ for cohort 1 and cohort 2 respectively. For each configuration, 9 classifying methods and their respective 50 runs were analyzed from which these percentages were calculated.

The trends in performance described above were the same for the balanced set-ups (Figs. 3, 4 and Supplementary Figs. S2 to S8 for cohort 1 and Supplementary Figs. S9 to S17 for cohort 2). The number of cases for which $Y_{i,C_m}^{VS} \pm SY_{i,C_m}^{VS}$ included $Y_{i,C_m}^{TES}$ (Table 2 for cohort 1 and Supplementary Table S3 for cohort 2) varied also similarly for balanced and unbalanced conditions.

Figure 3a and Supplementary Figs. S2a to S8a demonstrate that for cohort 1, all classifiers reached stability (unbalanced cases) for 160 patients (S4, $C_4$ and $C_9$), for 200 patients (S5, $C_8$), for 280 patients (S7, $C_1$ and $C_3$), for 320 patients (S8, $C_6$ and $C_7$) and for 360 patients (S9, $C_2$ and $C_5$). For cohort 2, all classifiers reached stability (Supplementary Figs. S9a to S17a) for 54 patients (S2, $C_1$ and $C_2$), 81 patients (S3, $C_3$, $C_5$, and $C_7$) and 108 patients (S4, $C_4$, $C_6$, $C_8$ and $C_9$). Arrows show from which set-up stability is observed.

When using all patients, we found that 7 out of 9 classifiers for cohort 1 and all classifiers for cohort 2 yielded a YI ≥ 0.70. AUC were equal to or greater than 0.75 for 7 out of 9 classifiers for cohort 1 and for all classifiers for cohort 2. For synthetic cohorts 1 and 2, when using all patients, the 5 classifiers led to YI between −0.10 and 0.10 and an AUC between 0.45 and 0.55.

**Prediction of the classification model performance - Univariate models.** For the univariate analyses, all features used in cohorts 1 and 2 reached stability but not necessarily with a predictive value as some features converged to a $Y_{ROC}^{TRS+VS}$ of 0. For each feature independently, QF1 $\leq$ QF2 for 76% of the 731 set-ups (17 experimental set-ups times 43 features). With as few as 36 patients, QF1 $\leq$ QF2 for 37 of the 43 features. For cohort 2, QF1 $\leq$ QF2 in 98% of the 570 set-ups (19 experimental set-ups times 30 features). With only 19 patients, QF1 $\leq$ QF2 for 30 of the 30 features. The predicted $Yec_{ROC}(N)$ were less than 0.20 for synthetic cohort 1 in 713/731 cases. For synthetic cohort 2, $Yec_{ROC}(N)$ was less than 0.20 in 507/570 cases.

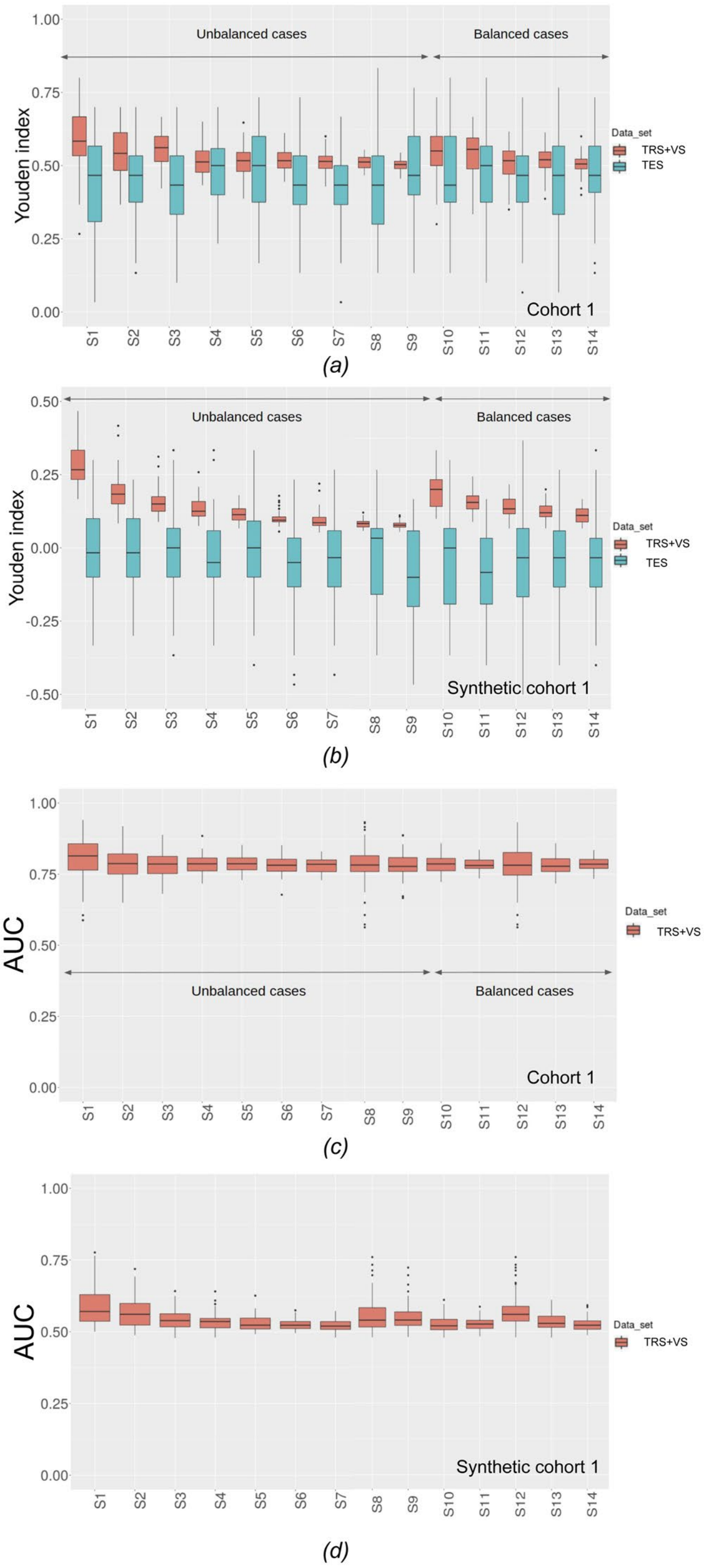

**Figure 2.** (**a**) $Y_{i,ROC}^{TRS+VS}$ (red) and $Y_{i,ROC}^{TES}$ (blue) for the 14 different experimental conditions of cohort 1 for the ROC study involving SUVmin. (**b**) Same as 2.a for the ROC study of synthetic cohort 1 with no predictive information involving SUVmin. (**c**) AUC on TRS + VS for the 14 different experimental conditions of cohort 1 for the ROC study involving SUVmin. (**d**) Same as 2.c for synthetic cohort 1 with no predictive information involving SUVmin.

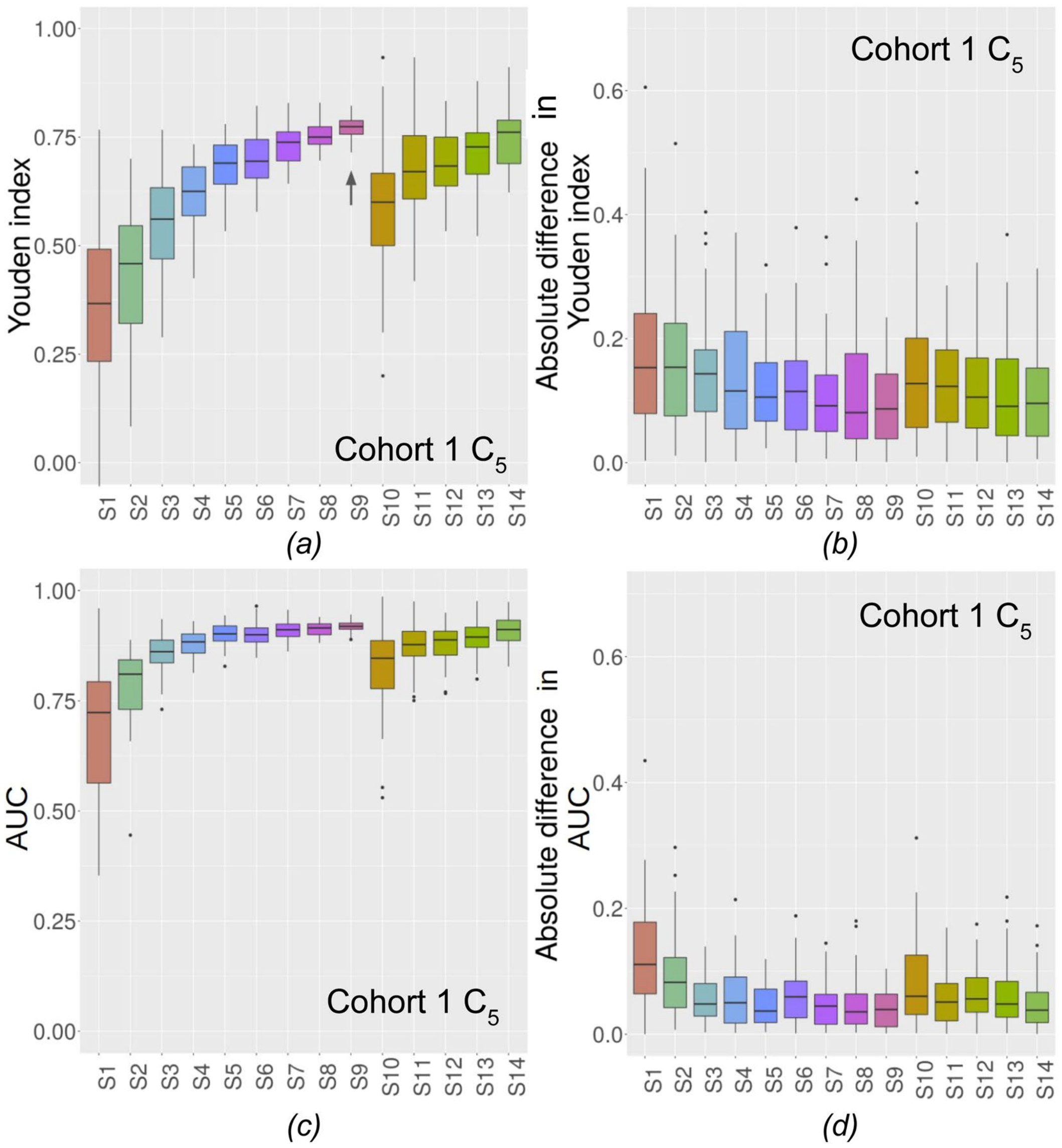

**Figure 3.** (**a**) $Y^{VS}_{C_5}$ for the 14 experimental conditions for $C_5$. ↑ shows when stability is reached for unbalanced cases (cohort 1). (**b**) Absolute differences in Youden index between $Y^{VS}_{C_5}$ and $Y^{TES}_{C_5}$ for $C_5$ (cohort 1). (**c**) AUC on VS for the 14 experimental conditions for $C_5$. (**d**) Absolute differences in AUC index between VS and TES for $C_5$ (cohort 1).

**Prediction of the classification model performance - Multivariate models.** The QF values for each experimental condition for ML classifiers are shown in Table 3 (YI) and Supplementary Table S4 (AUC) for cohort 1 and Supplementary Table S5 (YI) and Supplementary Table S6 (AUC) for cohort 2.

For cohort 1, for the Youden index, QF1′ was less than QF2′ in 183/216 cases. In terms of AUC, QF1′ was less than QF2′ in 184/216 cases. In addition, AUC was predicted within 5% of the AUC obtained when building the models using all patients in 78% of the cases. For cohort 2, QF1′ was less than QF2′ in 151/171 cases. In terms of AUC, QF1′ was less than QF2′ in 149/171 cases. In addition, AUC was predicted within 5% of the AUC obtained when building the models using all patients in 98% of the cases.

Regarding the synthetic datasets, $Yec_{C_m}(N)$ was always less than or equal to 0.10 for synthetic cohort 1 (except in 1/120 situation with predicted Youden index of 0.35) and for synthetic cohort 2 (except in 14/95 situations with predicted Youden index of 0.11 (2), 0.12 (3), 0.13, 0.14, 0.15 (3), 0.17, 0.22, 0.25, 0.47). For the predicted AUC, it was always less than or equal to 0.55 for synthetic cohort 1 (except in 5/120 situations with predicted AUC of 0.56 (2), 0.57 (2) and 0.66) and for synthetic cohort 2 (except in 14/95 situations with predicted AUC of 0.56 (5), 0.47 (2), 0.58 (1), 0.59 (2), 0.62, 0.65, 0.67 and 0.73).

**Use case.** Figure 5 illustrates the logarithmic fit used to determine $Yec_{C_9}(84)$ expected at stability from the 88 available patients. $Yec_{C_9}(84)$ obtained with the proposed downsampling approach was 0.82 against 0.54 when estimating $Y^{VS}_{C_9}(88)$ without any downsampling, while $Y^{VS}_{C_9}(414)$ is 0.81. The predicted AUC was 0.91, against 0.77 without downsampling and AUC observed using all 414 patients was 0.91.

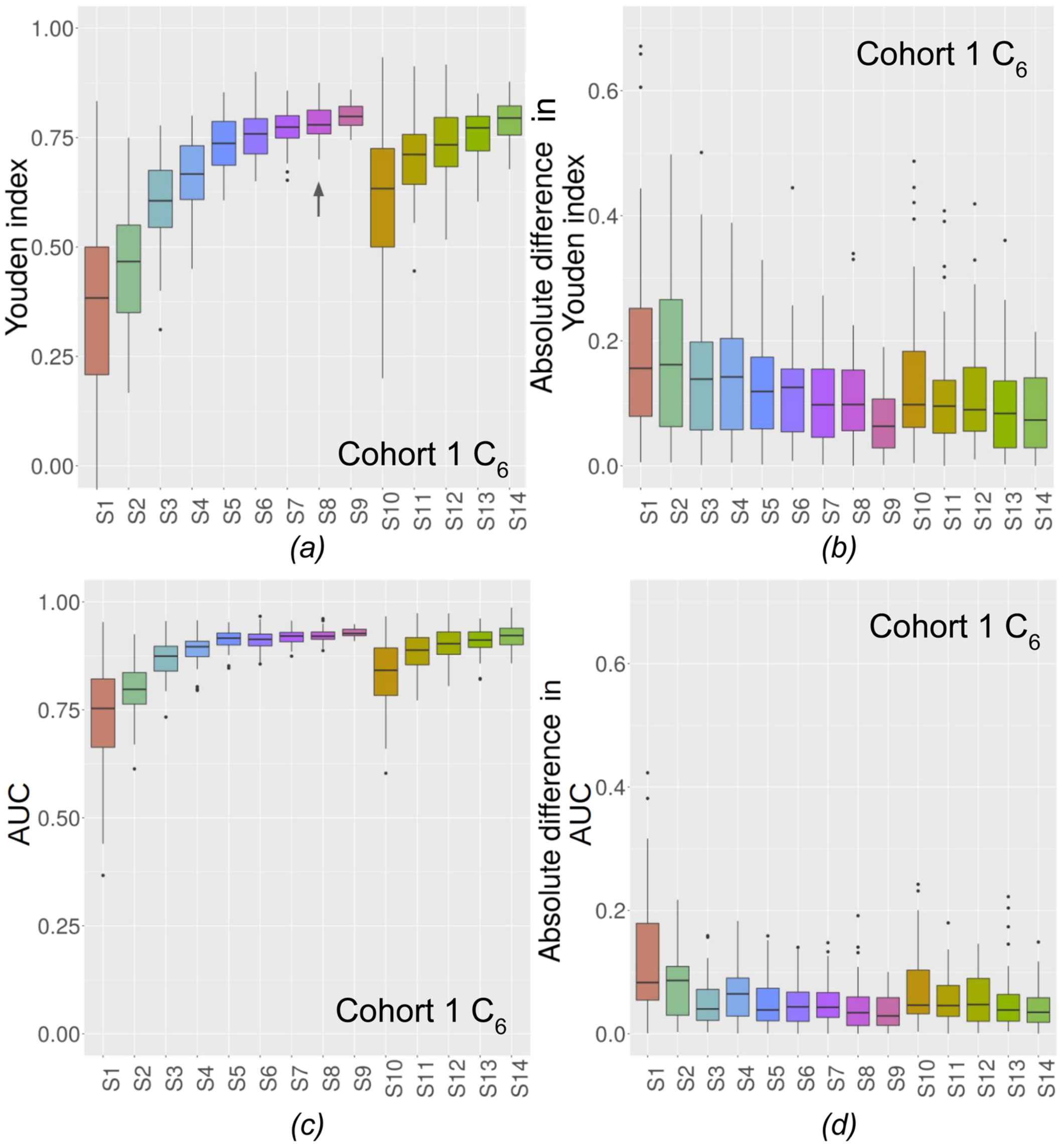

**Figure 4.** (**a**) $Y^{VS}_{C_6}$ for the 14 experimental conditions for $C_6$. ↑ shows when stability is reached for unbalanced cases (cohort 1). (**b**) Absolute differences in Youden index between $Y^{VS}_{C_6}$ and $Y^{TES}_{C_6}$ for $C_6$ (cohort 1). (**c**) AUC on VS for the 14 experimental conditions for $C_6$. (**d**) Absolute differences in AUC index between VS and TES for $C_6$ (cohort 1).

| Set-ups | S1 | S2 | S3 | S4 | S5 | S6 | S7 | S8 | S9 | S10 | S11 | S12 | S13 | S14 |
|---|---|---|---|---|---|---|---|---|---|---|---|---|---|---|
| $Y^{TES}_{i,C_m} \in Y^{VS}_{i,C_m} \pm SY^{VS}_{i,C_m}$ | 47 | 40 | 59 | 59 | 62 | 62 | 67 | 64 | 74 | 53 | 64 | 66 | 65 | 74 |

**Table 2.** For each set-up, percentage of the 450 cases for which $Y^{TES}_{i,C_m}$ was included in $Y^{VS}_{i,C_m} \pm SY^{VS}_{i,C_m}$ (cohort 1).

## Discussion

In radiomic studies, an important step is to determine whether the set of radiomic features contains relevant information for a given classification task given the available data. We designed a downsampling method to answer that question by accounting for the experimental conditions. To do so, we started from data for which the presence of predictive features was demonstrated and studied the behavior of different classifiers when degrading the experimental conditions to identify trends. The impact of having balanced or unbalanced dataset was investigated, as well as the role of feature selection. The proposed downsampling approach uses the observed trends to estimate the classification performance to be expected when enough patients are available.

| Number of patients | LR | | RFELR | | PCALR | | ROCLR | | SVM | | RFESVM | | PCASVM | | ROCSVM | | LASSO | |
|---|---|---|---|---|---|---|---|---|---|---|---|---|---|---|---|---|---|---|
| | QF1′ | QF2′ | QF1′ | QF2′ | QF1′ | QF2′ | QF1′ | QF2′ | QF1′ | QF2′ | QF1′ | QF2′ | QF1′ | QF2′ | QF1′ | QF2′ | QF1′ | QF2′ |
| 220 | 0.03 | 0.09 | 0.04 | 0.07 | **0.06** | 0.05 | **0.1** | 0.02 | 0.02 | 0.11 | 0.05 | 0.09 | **0.06** | 0.04 | **0.1** | 0.02 | 0.04 | 0.06 |
| 212 | 0.04 | 0.09 | 0.01 | 0.09 | 0.02 | 0.05 | **0.09** | 0.01 | 0.02 | 0.13 | 0.04 | 0.09 | **0.07** | 0.05 | **0.08** | 0.02 | 0.04 | 0.07 |
| 204 | 0.05 | 0.08 | 0.03 | 0.08 | **0.05** | 0.04 | **0.13** | 0.00 | 0.02 | 0.14 | 0.00 | 0.1 | **0.08** | 0.05 | **0.08** | 0.03 | 0.07 | 0.07 |
| 196 | 0.05 | 0.08 | 0.02 | 0.1 | 0.04 | 0.04 | **0.12** | 0.02 | 0.00 | 0.14 | 0.01 | 0.11 | **0.07** | 0.04 | **0.13** | 0.04 | 0.08 | 0.09 |
| 188 | 0.04 | 0.08 | 0.00 | 0.11 | 0.04 | 0.05 | **0.13** | 0.01 | 0.03 | 0.15 | 0.01 | 0.11 | **0.12** | 0.06 | **0.13** | 0.02 | 0.04 | 0.06 |
| 180 | 0.08 | 0.09 | 0.00 | 0.12 | 0.03 | 0.06 | **0.08** | 0.02 | 0.02 | 0.14 | 0.02 | 0.12 | 0.04 | 0.05 | **0.06** | 0.05 | 0.08 | 0.08 |
| 172 | 0.05 | 0.1 | 0.05 | 0.09 | 0.03 | 0.07 | **0.14** | 0.01 | 0.06 | 0.14 | 0.04 | 0.11 | **0.1** | 0.06 | **0.13** | 0.05 | 0.07 | 0.08 |
| 164 | **0.1** | 0.09 | 0.04 | 0.14 | 0.02 | 0.09 | **0.11** | 0.03 | 0.01 | 0.15 | 0.03 | 0.14 | **0.11** | 0.06 | 0.05 | 0.05 | 0.05 | 0.1 |
| 156 | 0.01 | 0.12 | 0.03 | 0.14 | 0.01 | 0.09 | **0.12** | 0.03 | 0.01 | 0.18 | 0.00 | 0.18 | 0.06 | 0.09 | 0.05 | 0.09 | 0.03 | 0.1 |
| 148 | 0.03 | 0.12 | 0.02 | 0.16 | 0.01 | 0.11 | **0.09** | 0.04 | 0.02 | 0.2 | 0.09 | 0.17 | 0.08 | 0.1 | 0.04 | 0.08 | 0.05 | 0.13 |
| 140 | 0.03 | 0.12 | 0.01 | 0.16 | 0.05 | 0.13 | **0.14** | 0.04 | 0.08 | 0.24 | 0.09 | 0.19 | **0.1** | 0.08 | 0.00 | 0.08 | 0.03 | 0.13 |
| 132 | 0.05 | 0.12 | 0.05 | 0.19 | 0.04 | 0.12 | 0.09 | 0.11 | 0.08 | 0.22 | 0.04 | 0.17 | 0.00 | 0.13 | 0.04 | 0.13 | **0.17** | 0.11 |
| 124 | 0.05 | 0.11 | 0.05 | 0.18 | 0.06 | 0.13 | **0.09** | 0.03 | 0.00 | 0.22 | 0.11 | 0.22 | 0.06 | 0.12 | 0.01 | 0.11 | 0.05 | 0.14 |
| 116 | 0.00 | 0.22 | 0.06 | 0.26 | 0.03 | 0.18 | 0.02 | 0.09 | 0.12 | 0.27 | 0.1 | 0.26 | 0.05 | 0.17 | 0.05 | 0.15 | 0.02 | 0.19 |
| 108 | 0.06 | 0.23 | 0.03 | 0.27 | 0.07 | 0.22 | 0.06 | 0.13 | 0.1 | 0.25 | 0.01 | 0.26 | 0.04 | 0.17 | 0.1 | 0.18 | 0.08 | 0.21 |
| 100 | 0.02 | 0.22 | 0.00 | 0.26 | 0.07 | 0.19 | 0.03 | 0.14 | 0.06 | 0.29 | 0.00 | 0.28 | **0.29** | 0.17 | 0.01 | 0.21 | 0.01 | 0.19 |
| 92 | 0.04 | 0.21 | 0.11 | 0.28 | 0.00 | 0.25 | 0.01 | 0.16 | 0.12 | 0.3 | 0.01 | 0.32 | 0.07 | 0.18 | 0.07 | 0.27 | 0.02 | 0.2 |
| 84 | 0.04 | 0.23 | 0.02 | 0.3 | 0.00 | 0.22 | 0.05 | 0.2 | 0.13 | 0.3 | 0.18 | 0.29 | 0.03 | 0.19 | 0.04 | 0.3 | 0.02 | 0.23 |
| 76 | 0.12 | 0.22 | 0.00 | 0.33 | 0.08 | 0.2 | 0.14 | 0.19 | 0.09 | 0.32 | 0.00 | 0.34 | 0.08 | 0.25 | 0.03 | 0.31 | 0.01 | 0.29 |
| 68 | 0.09 | 0.28 | 0.01 | 0.38 | 0.05 | 0.24 | 0.09 | 0.19 | 0.03 | 0.34 | 0.00 | 0.37 | 0.04 | 0.24 | 0.21 | 0.33 | 0.03 | 0.29 |
| 60 | 0.05 | 0.26 | 0.01 | 0.31 | 0.05 | 0.27 | 0.16 | 0.23 | 0.2 | 0.32 | 0.26 | 0.39 | 0.08 | 0.28 | 0.15 | 0.31 | 0.1 | 0.33 |
| 52 | 0.12 | 0.31 | 0.06 | 0.36 | 0.12 | 0.29 | 0.01 | 0.22 | 0.04 | 0.37 | 0.05 | 0.36 | 0.14 | 0.32 | 0.26 | 0.3 | 0.08 | 0.34 |
| 44 | 0.22 | 0.33 | 0.09 | 0.39 | 0.03 | 0.34 | 0.07 | 0.3 | 0.19 | 0.4 | 0.18 | 0.44 | 0.08 | 0.31 | 0.05 | 0.39 | 0.07 | 0.38 |
| 36 | **0.44** | 0.41 | 0.01 | 0.45 | 0.22 | 0.36 | 0.12 | 0.37 | 0.11 | 0.52 | 0.04 | 0.47 | 0.07 | 0.39 | 0.36 | 0.43 | 0.03 | 0.41 |

**Table 3.** QF in each experimental condition for each classifier that ultimately reaches stability (cohort 1). Bold values: QF1′ ≥ QF2′.

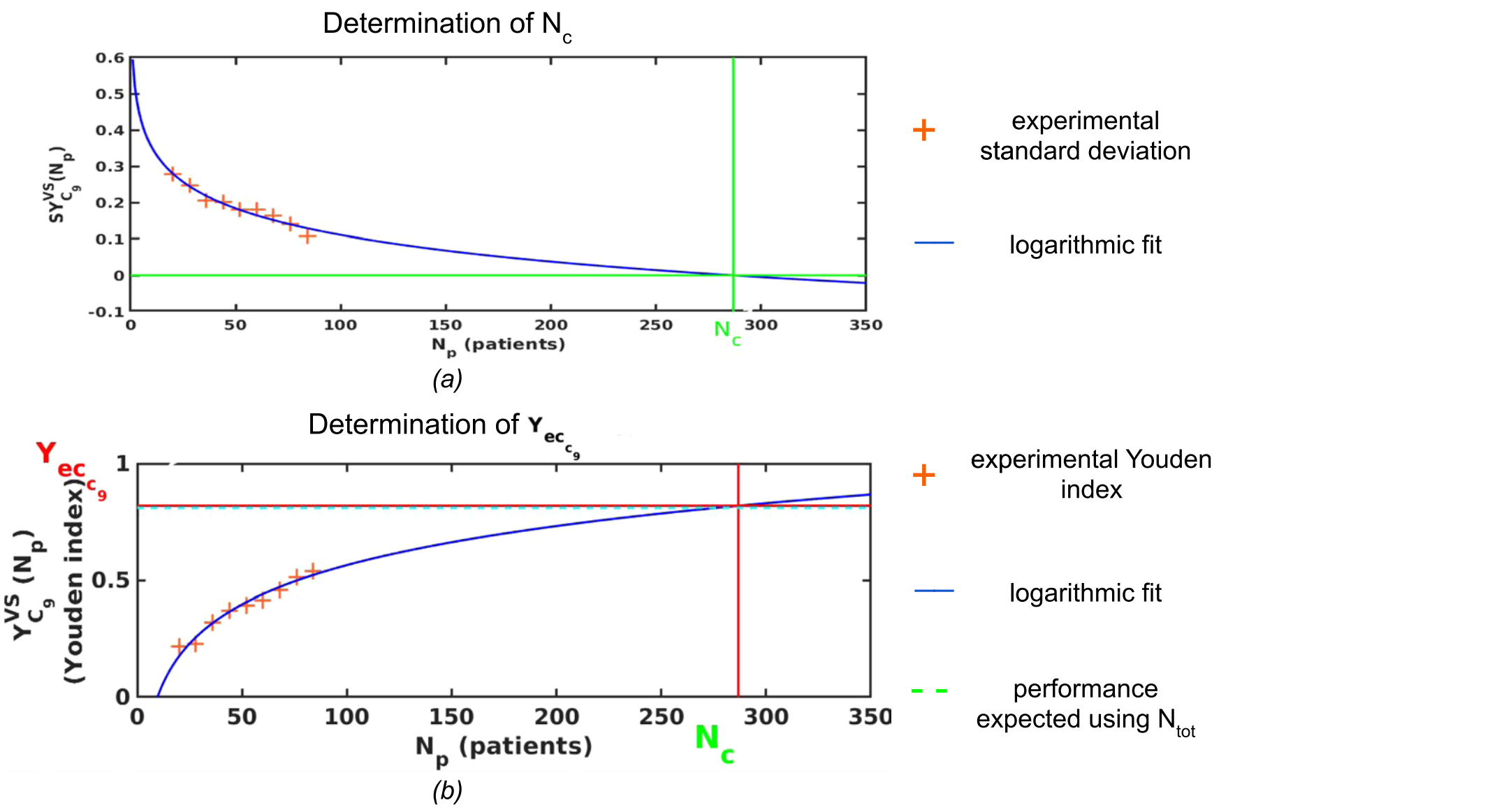

**Figure 5.** (**a**) Use case: fitting of a logarithmic curve (Eq. 3) with each subsample's standard deviation to find $N_c$. (**b**) Use case: fitting of the logarithmic curve of Eq. 4 with each subsample's Youden index, assuming that 84 patients are available and model is built using $C_9$.

For our study, we used two large public datasets for which features were provided together with results of predictive models, so as to be able to compare our results with those published independently by other investigators. Cohort 1 was a radiomic data set while cohort 2 corresponded to image-based features derived from FNA images (not genuine radiomic data as FNA are not radiology images). This latter cohort was used because of our difficulty in identifying large enough cohorts for which modelling results would not improve significantly, on average, if adding more patients, as demonstrated in Fig. 3a, 4a and Supplementary Figs. S2 to S17.

We investigated univariate radiomic models based on ROC analysis as this approach is still frequently used in the context of radiomics[10]. In ROC analyses, Fig. 2a shows that when the number of patients increases, $Y_{ROC}^{TRS+VS}$ decreases as well as the associated uncertainty. This has not been reported before in the context of radiomics, to the best of our knowledge, but the same trend was observed in a completely different context[11]. From a practical point of view, this suggests that when the number of patients is insufficient, univariate analysis is likely to overestimate the actual classification performance as expressed by the Youden index that could be observed with a high number of patients. Yet, the AUC does not change much with the number of patients, making it more reliable than the Youden index when the cohort used to assess the predictive value of a radiomic feature is small. In addition and as expected, the difference between $Y_{ROC}^{TRS+VS}$ and $Y_{ROC}^{TES}$ observed in the test set decreases when the number of patients increases, with a less severe overestimation of $Y_{ROC}^{TRS+VS}$ compared to the one observed when performance reached stability. $Y_{ROC}^{TRS+VS}$ became non-significantly different from $Y_{ROC}^{TES}$ observed when at least 200 patients (150PT and 50MT) were used for cohort 1 (189 patients for cohort 2–70MM and 119BM). When the conditions are poor (40 patients with 30 PT and 10 MT), up to 80% of the features identified as predictive on the TRS + VS did actually not have any predictive power in the test dataset for cohort 1. For cohort 2, a similar trend was observed with up to 64% of the features that were not confirmed as being predictive in the test set in the strongly degraded set-up (27 patients - 10MM and17BM). The minimum number of patients needed to properly identify the predictive features depends on the data set as it depends on the predictive power of each feature.

In multivariate models, unlike for the univariate analysis, $Y_{C_m}^{VS}$ increases when the experimental situation becomes more favorable (more cases, balanced samples) (Figs. 3a and 4a) and gets close to $Y_{C_m}^{TES}$ when the experimental conditions improve. This means that with an insufficient number of patients, multivariate models will actually often underestimate the classification performance that could be reached if the model was built with a large enough number of patients. AUC shows the same behaviour and this trend is the same whatever the classifier, and whether feature selection is used or not. Depending on the dataset and on the classifier, the univariate or multivariate analysis can converge faster. For instance, with cohort 1, univariate analysis $Y_{ROC}^{TRS+VS}$ converged faster (200 patients) than ML classifier $Y_{C_m}^{VS}$ (160 to 360 patients) for most features. Yet, when stability was reached, $Y_{C_m}^{VS}$ and AUC were higher with multivariate techniques than $Y_{ROC}^{TRS+VS}$. The inverse trend for convergence was observed for cohort 2. Indeed, ML classifiers reached stability faster (51 to 125 patients) but with still higher $Y_{C_m}^{VS}$ than univariate analysis (162). For multivariate methods, both the $Y_{C_m}^{VS}$ and AUC obtained at stability and the number of patients needed to reach stability depend on the classifier.

The classifier performance $Y_{C_m}^{TES}$ were statistically different between prediction models but were all within 0.10 Youden unit (0.10 AUC unit), suggesting that one can conclude at the predictive value of features even if the classifier that is chosen is not the most appropriate one.

Regarding feature selection, we compared selecting features from univariate ROC analysis with multivariate feature selection (RFE, PCA, LASSO) when used with multivariate methods (LR and SVM). There is no clear trend considering the performance of these two types of approaches for feature selection: whether one approach outperforms the other depends on the dataset. The selected features were more similar between two ML methods involving RFE (eg. $C_4$ and $C_7$) than between $ROC_{fr}$ and RFE-based ML (results not shown). The most efficient feature selection technique depended on the ML algorithm subsequently used: in cohort 1, RFE was the most efficient technique when either associated with LR or SVM. Both were outperformed by LASSO, and this was the case for most configurations in cohort 1. In cohort 2, RFE was the most efficient technique when associated with LR and PCA was the most efficient one when associated to SVM. This variability is consistent with what has been reported in the literature[12] (Random Forest)[13], (weighted k-nearest neighborhood)[14], (Fisher score + k-nearest neighborhood)[15], (RFE-SVM)[16], (LDA).

Table 2 shows that when the number of patients is not sufficient (less than 120 patients - 90PT and 30MT for cohort 1), $Y_{C_m}^{VS}$ as estimated from the cross-validation approach is not reliable. For instance, in cohort 1 with 40 patients, only 47% of the $Y_{i,C_m}^{TES}$ measured using cross-validation were within $Y_{i,C_m}^{VS} \pm SY_{i,C_m}^{VS}$, while for 360 patients, this percentage was 74%. In balanced situations (ie. 60 PL and 60 MT), the estimation was more reliable than for the unbalanced cases with the same number of patients (90 PL and 30 MT). Still, in order to ensure that predictions are reliable, an external dataset is absolutely needed as already suggested[17].

All results (Figs. 3, 4 and Supplementary Figs. S2 to S17) comparing balanced set-up (same number of patients in the two groups) with a set up with more than 3 times more patients in one group compared to the other suggest that using a larger number of patients is preferable, even if the two resulting groups are unbalanced.

To predict the performance of the model that can be reached if enough patients were available, we described an empirical downsampling approach. Downsampling the number of patients made it possible to determine how the Youden index (ROC and ML) and the AUC (ML) evolve as a function of the number of patients and to extrapolate the expected value if a sufficient number of patients was available. Using this approach, for ML techniques, $Yec_{C_m}(N)$ was almost always (85% of the 216 set-ups for cohort 1 and 87% of the 171 set-ups for cohort 2) closer to $Y_{C_m}^{VS}(N_{tot})$ obtained when stability is reached than $Y_{C_m}^{VS}(N)$ directly obtained using all available patients without downsampling. Conclusions are the same for the AUC predictions. In the synthetic PET datasets, where no predictive model was expected, the downsampling method almost always concluded at the absence of information for ROC univariate analysis and ML techniques. $Yec_{C_m}(N)$ was less than or equal to 0.10 for both synthetic cohorts and the AUC were less than or equal to 0.55, demonstrating the absence of predictive factors in the data and validating the specificity of the approach.

In the studied cohorts, we found that at least 80 patients (40 in each of the 2 groups to be distinguished) were needed to use this downsampling approach and estimate well $Yec_{C_m}(N)$ and AUC (within 5%) of the one observed with the largest number of patients. This was almost always less patients than what would be needed without the downsampling method. Therefore, for a given number of patients, the downsampling method provides a better estimate of the model performance than an estimate based only on the available patients without any downsampling.

The proposed downsampling method has limitations. In particular, a minimum number of patients is required to get reliable estimates of the performance of the model. A minimum of 40 patients for each status is recommended. Another limitation is that the downsampling method estimates classification performance to be expected when designing a model from a large enough sample of data but does not give clear indications about which ML method might be the most appropriate. Therefore, it should be seen as a screening method, to help make a go/no go decision for further model design.

Last, our study was conducted on two representative datasets, but further validation using other datasets should be performed.

In conclusion, we reported how the radiomic model performance estimated from datasets including an insufficient number of patients might be biased. We also described an original downsampling approach that yields a better estimate of the model performance characterized by the Youden index or the AUC expected when enough patients are available than the performance directly estimated from the available limited dataset.

## References

1. Gillies, R. J., Kinahan, P. E. & Hricak, H. Radiomics: Images are more than pictures, they are data. *Radiology* **278**, 563–577 (2015).
2. Li, H., Galperin-Aizenberg, M., Pryma, D., Simone, C. B. & Fan, Y. Unsupervised machine learning of radiomic features for predicting treatment response and overall survival of early stage non-small cell lung cancer patients treated with stereotactic body radiation therapy. *Radiother Oncol.* **129**, 218–226 (2018).
3. Leger, S. *et al.* A comparative study of machine learning methods for time-to-event survival data for radiomics risk modelling. *Sci Reports* **7**, 13206 (2017).
4. Reuzé, S. *et al.* Prediction of cervical cancer recurrence using textural features extracted from 18 F-FDG PET images acquired with different scanners. *Oncotarget* **8**, 43169–43179 (2017).
5. Figueroa, R. L., Zeng-Treitler, Q., Kandula, S. & Ngo, L. H. Predicting sample size required for classification performance. *BMC Med Inform Decis Mak.* **12**, 8 (2012).
6. Kirienko, M. *et al.* Ability of FDG PET and CT radiomics features to differentiate between primary and metastatic lung lesions. *Eur J Nucl Med Mol Imaging.* **45**, 1649–1660 (2018).
7. Nioche, C. *et al.* LIFEx: a freeware for radiomic feature calculation in multimodality imaging to accelerate advances in the characterization of tumor heterogeneity. *Cancer Res.* **78**, 4786–4789 (2018).
8. Wolberg, W. H., Street, W. N., Heisey, D. M. & Mangasarian, O. L. Computer-derived nuclear features distinguish malignant from benign breast cytology. *Human Patholology.* **26**, 792–796 (1995).
9. Pedregosa, F. *et al.* Scikit-learn: Machine learning in Python. *JMLR.* **12**, 2825–2830 (2011).
10. Gill, T. S. *et al.* Juxtatumoral perinephric fat analysis in clear cell renal cell carcinoma. *Abdom Radiol.* **44**, 1470–1480 (2018).
11. McPherson, J. M., Jetz, W. & Rogers, D. J. The effects of species' range sizes on the accuracy of distribution models: ecological phenomenon or statistical artefact? *J Appl Ecol.* **41**, 811–823 (2004).
12. Kang, C., Huo, Y., Xin, L., Tian, B. & Yu, B. Dual-energy CT texture analysis with machine learning for the evaluation and characterization of cervical lymphadenopathy. *Comput Struct Biotechnol J.* **17**, 1009–1015 (2019).
13. Acar, E. *et al.* Machine learning for differentiating metastatic and completely responded sclerotic bone lesion in prostate cancer: a retrospective radiomics study. *BJR* **92**, 20190286, https://doi.org/10.1259/bjr.20190286 (2019).
14. Du, D. et al. Machine learning methods for optimal radiomics-based differentiation between recurrence and inflammation: application to nasopharyngeal carcinoma post-therapy PET/CT image. *Mol Imaging Biol*. https://doi.org/10.1007/s11307-019-01411-9 (2019).
15. Zhang, Y. *et al.* Radiomics analysis for the differentiation of autoimmune pancreatitis and pancreatic ductal adenocarcinoma in $^{18}$F-FDG PET/CT. *Med Phys.* https://doi.org/10.1002/mp.13733 (2019).
16. Xie, T. *et al.* Machine learning-based analysis of MR multiparametric radiomics for the subtype classification of breast cancer. *Front Oncol.* **9**, 505 (2019).
17. Reuzé, S. *et al.* Radiomics in nuclear medicine applied to radiation therapy: methods, pitfalls, and challenges. *Int J Radiat Oncol Biol Phys.* **102**, 1117–1142 (2018).

## Acknowledgements

The data set of cohort 2 is the Wisconsin Breast Cancer data set created by W.H. Wolberg, General Surgery Department, University of Wisconsin, Clinical Sciences Center, W.N. Street and O.L. Mangasarian, Computer Sciences Department, University of Wisconsin. The repository's reference is: Dua, D. and Graff, C. (2019). UCI Machine Learning Repository [http://archive.ics.uci.edu/ml]. Irvine, CA: University of California, School of Information and Computer Science.

## Author contributions

A.-S.D., F.F. and I.B. contributed to the study design. A.-S.D. performed the statistical analysis (ROC. analysis, machine learning and downsampling method). A.-S.D. and I.B. wrote the main manuscript. All authors reviewed the manuscript.

## Competing interests

The authors declare no competing interests.

## Additional information

**Correspondence** and requests for materials should be addressed to A.-S.D.